\documentclass[3p,times,twocolumn]{elsarticle}

\usepackage{ecrc}

\volume{00}

\firstpage{1}

\journalname{Nuclear Physics B Proceedings Supplement}

\runauth{biii}

\jid{nuphbp}

\jnltitlelogo{Nuclear Physics B Proceedings Supplement}

\usepackage{amssymb}

\newcommand{\Eq}[1]{{Eq.~({\ref{#1}})}}

\newcommand{\bea}{\begin{eqnarray}}
\newcommand{\eea}{\end{eqnarray}}
\newcommand{\beq}{\begin{equation}}
\newcommand{\eeq}{\end{equation}}
\newcommand{\beas}{\begin{eqnarray*}}
\newcommand{\eeas}{\end{eqnarray*}}

\def\p{{\bf p}}

\def\x{{\bf x}}

\begin{document}

\begin{frontmatter}



\dochead{}

\title{Inhomogeneous phases and chiral symmetry breaking}

\author{Stefano Carignano, Efrain J. Ferrer, and Vivian de la Incera}

\address{University of Texas at El Paso, Department of Physics, 500 W University Ave., El Paso, TX 79968}

\begin{abstract}
A significant fraction of the current efforts in the QCD research community is focused on characterizing the phases of strong-interaction matter that occur at finite densities and temperatures.
  So far, most of the experimental probes have been limited to the relatively narrow window of the QCD phase diagram characterized by high temperatures and low chemical potentials, explored in high-energy ion collision experiments at RHIC and LHC. 
  More recently, some new insight on the finite chemical potential region has been obtained by the energy-beam scan program at RHIC, aimed at possibly determining the existence of a critical point in the QCD phase transition.
 On the theoretical side, studies of strong interactions are also limited by the reliability of available methods. 
 While the zero density, finite temperature region or the zero temperature, superdense region can be investigated with the help of well-established approaches like lattice and weakly coupled QCD respectively, the study of the intermediate densities and temperatures region has to rely on effective models and nonperturbative methods, some of which are still being developed.
  In the past few years, a growing number of compelling arguments, backed up by model calculations, pointed out that the intermediate-density region of the QCD phase diagram may be characterized by the formation of inhomogeneous condensates which spontaneously break some of the spatial symmetries of the theory.  In the following we provide a brief recapitulation of these arguments and describe some recent results in a 3+1-dimensional QCD-inspired NJL model with quark-hole condensation in the form of a plane wave in the scalar and tensor channels. This model exhibits particular features in close analogy to its 1+1-dimensional counterpart, most notably an asymmetric spectral density and the arising of an anomalous contribution to the free energy.
\end{abstract}


\end{frontmatter}


\section{The Case for Inhomogeneous Phases}
\label{S1}
Mapping all the phases of QCD in the temperature-density plane is a goal intensely sought after by many theoretical and experimental efforts \cite{QCDreviews}.
 Thanks to the asymptotic freedom properties of the theory, the extreme regions of the QCD phase diagram are weakly coupled and hence well understood; 
 they are the quark-gluon plasma in the high-temperature/low-density corner and the Color-Flavor-Locked (CFL) superconducting phase \cite{alf-raj-wil-99/537} on the opposite side.  At low temperatures and densities quarks are confined inside hadrons, whose interactions can be phenomenologically described by conventional nuclear physics.

Somewhere in the region of intermediate temperatures and densities, one expects two kinds of phase transitions to occur, the first related to deconfinement and liberations of quark degrees of freedom from hadrons, the second associated to the restoration of chiral symmetry, which is spontaneously broken in vacuum. 
Indeed, it is known that 
 at low temperatures and densities, quarks acquire a large constituent mass due to the formation of a (spatially homogeneous) chiral condensate as a result of quarks-antiquark pairing. With increasing density, this condensate becomes disfavored due to the high energy cost (at least twice the Fermi energy) required to excite the antiquarks from the Dirac sea to the Fermi surface. What phase forms when this ``traditional'' vacuum chiral condensate disappears is not clear yet. 
 It has been argued that at sufficiently high densities,
 co-moving quarks and holes at the Fermi surface
 may pair through a mechanism that is analogous to the Overhauser one \cite{Overhauser}, giving rise to a spatially modulated condensate \cite{largeNQCD}.
 Since the intermediate and high density regions are expected to be within 
 the realm of color-superconductivity (CS) as well, 
  a competition is expected to develop between the Bardeen-Cooper-Schrieffer (BCS) mechanism \cite{CS} pairing quarks with quarks of opposite momenta, and this Overhauser-type inhomogeneous pairing 
  \cite{large_finiteN,PRD62,PRD63}.

Notice that the BCS pairing mechanism takes place because in a dense and cold system of (approximately) massless quarks 
it costs no energy to excite unpaired quarks at the Fermi surface.
 Any attractive interaction, no matter how weak, will then favor pairing, thereby decreasing the system's energy through condensation. An attractive quark-quark interaction is embedded in the fundamental interaction of QCD, and the stronger channel would pair quarks at the opposite sites of the Fermi surface. This is the reason why color superconductivity should naturally occur at high densities. 

BCS pairing is clearly favored in the weak coupling regime of highly dense QCD because the BCS gap gets contributions from the entire Fermi surface, and the density of states participating in the diquark formation increases with the Fermi surface radius. However, at intermediate densities this mechanism suffers from a pairing stress arising due to the difference in the chemical potentials of different flavors. Under these circumstances, Cooper pairing with opposite momenta becomes less favorable because quarks cannot pair at their respective Fermi surfaces, requiring an additional energy cost. Still, BCS pairing survives as long as the gain in condensation energy remains larger than the pairing stress. Once the stress becomes too large, the system acquires chromomagnetic instabilities \cite{ChMInst}, indicating that one is working in the wrong ground state. A possible solution is BCS pairing with a net momentum or other forms of inhomogeneous color-superconducting ground states, but despite a variety of propositions to cure the instability \cite{IHCS}, the question of which ground CS state is energetically favored still remains open.

In contrast to the BCS mechanism, which uniformly covers the entire Fermi surface, the Overhauser pairing takes place in patches of the Fermi surface and requires a sufficiently strong coupling to occur in (3+1) dimensions,
at least for the physical case of three colors ($N_c=3$) \cite{PRD63}. On the other hand, the BCS issue of different flavor chemical potentials at intermediate densities does not affect the Overhauser pairing because it is a flavor singlet.

The mechanism behind Overhauser pairing is based on the nesting between the Fermi surfaces of particles and holes. It occurs when segments of these surfaces are approximately parallel to each other and thus can be connected by a common nesting vector. In one spatial dimension, the nesting is automatic and complete because the two Fermi "surfaces" reduce to two points. In three spatial dimensions however, it requires deformation of the Fermi surface and can occur only within limited regions (or ``patches''). For large chemical potentials the Fermi surface in a (3+1)-dimensional theory tends to "flatten" and one expects a stronger nesting effect. 
 However,  the realization of Overhauser pairing in QCD requires a delicate balance among the different factors influencing the pairing mechanism. For instance, even though nesting is favored in a bigger Fermi surface, if the chemical potential is too large, the coupling becomes weaker and the BCS pairing may end up winning over the Overhauser one. The number of patches participating in the pairing phenomenon is also a factor in this balance of tendencies. The number of patches increases slowly with the density, but more adjacent patches with a smaller relative angle means more interaction between them, which in turn produces an energy cost that dominates at sufficiently large density \cite{PRD82}. Then, quark-hole pairing is expected to occur at densities large enough for the system to be in a quark phase ($\mu> \Lambda_{QCD}$), but small enough to support strongly coupled interactions and to keep patch interaction under control. It is in this region that Overhauser pairing has a real chance to win over BCS pairing. 
 
Understanding the fundamental physics involved and the phases that characterize the regime of densities between hadronic matter and CFL quark matter in the QCD phase diagram remains an outstanding and highly nontrivial task. In this context, looking for physically meaningful models exhibiting Overhauser pairing in the intermediate-density region is an important topic of investigation, one that promises to shed new light on an old challenging problem and is therefore gaining more and more interest.

\section{Effective QCD Models with Quark-Hole Pairing}
\label{S2}
There have been several attempts to model Overhauser pairing in QCD.  All the studies performed so far seem to indicate that (at least for not excessively high densities) one-dimensional spatially modulated chiral condensates are energetically preferred over higher-dimensional ones \cite{car12,abuki,interwievingCS}. In the following we shall therefore focus our discussion on inhomogeneous condensates which vary in only one spatial direction.

The idea that quark-hole pairing could be relevant for dense QCD was first discussed in the pioneering work of Deryagin, Grigoriev and Rubakov (DGR) \cite{largeNQCD}, who considered QCD in the large $N_c$ limit, using the ladder approximation and integrating out gluonic degrees of freedom in order to obtain a fermionic action with an effective vertex.
%
%
%
Considering the perturbative regime $g^2N_c \ll1$, these authors found a singularity in the effective four-fermion interaction of this model, triggering an instability
at the Fermi surface with respect to quark-hole pairing
 that leads to the formation of an inhomogeneous chiral condensate, proposed as
 \begin{equation}
\hspace{-.4cm} \Sigma(x,y)=2\cos[P_\mu (\frac{x_\mu+y_\mu}{2})] \int \frac{d^4q}{(2\pi)^4}e^{-iq(x-y)}F(q) \,,\;
\end{equation}
where $|P|=2P_F$ ($P_F$ being the Fermi momentum) and $F(q)$ is assumed to be nonzero only at relative small values of $q$. 

The DGR instability was obtained ignoring the screening effects from the fermion loops, thus requiring $N_c \rightarrow \infty$. A next development came in Ref. \cite{large_finiteN}, which explored whether the instability would be still operative at some finite $N_c$ and found using renormalization group arguments that, at least within a perturbative regime, the DGR instability occurs only at extremely large  $N_c$, $N_c\gtrsim 1000N_f$.
This picture might of course be dramatically modified when going away from the perturbative regime. Such an approach (still in the large $N_c$ limit) has been pursued in 
%
Ref. \cite{q-chiralspirals}, where the QCD quark-hole pairing phenomenon was revisited  in the context of quarkyonic matter \cite{quarkyonic-matter}. Quarkyonic matter has been argued to describe a new state of QCD at low temperatures and baryon densities large compared to the QCD scale, so that the Fermi sea is best described in terms of quark degrees of freedom, while excitations near the Fermi surface are color-confined mesons and baryons. Even though the arguments for the existence of this exotic matter are rigorous only for a large number of colors, this may not be a bad approximation for some range of densities at $N_c=3$ \cite{interwievingCS}.

A key observation in \cite{q-chiralspirals} was that in the large $N_c$ limit the gluon propagator is unaffected by quarks and hence it is the same as the confined vacuum propagator. Then, to investigate the gap equation these authors used the following gluon propagator
\begin{equation} \label{confD}
D_{00}(\omega,\mathbf{q})=\frac{\sigma}{|{\mathbf{q}^2}|^2}, \qquad D_{ij}(\omega,\mathbf{q})=\frac{\delta_{ij}-q_i q_j/q^2}{\omega^2 +\mathbf{q}^2},
\end{equation}
which is valid in the Coulomb gauge $\partial_i A_i=0$  at small momenta $|\mathbf{q}|<\Lambda_{QCD}$. Notice that the main channel of interaction comes from the non-perturbative confining part of the potential, determined by the $(0,0)$ component of the gluon propagator. 

Using (\ref{confD}), the Schwinger-Dyson equation for the fermion self-energy can be reduced to that of (1+1)-dimensional QCD at finite density, with ground state solution given by a plane wave generated by the sum of two chiral condensates, a so-called ``quarkyonic chiral spiral'', for which 
\beq
\label{sp}
\hspace{-.2cm}\langle\bar{\psi}\psi\rangle= \Delta \cos (2\mu x) \;, \quad
\langle\bar{\psi}\gamma_0\gamma_3\psi\rangle= \Delta \sin (2\mu x)  \,.
\eeq 

While the large $N_c$ limit has proven to be a helpful tool for understanding many properties of QCD, a complementary investigation of the more realistic three-color case would of course be highly desirable.

\section{NJL Models with Quark-Hole Pairing}
\label{S3}
Another line of investigation in the quest to model Overhauser pairing in QCD has been based on proposing four-fermion interaction models \`a la Nambu-Jona-Lasinio (NJL) \cite{NJL}. These models exhibit the same chiral symmetry as QCD and can be suitably extended to include the possible formation of inhomogeneous condensates, without being restricted to small coupling approximations (for a recent review on model results on inhomogeneous phases we refer the interested reader to Ref.\cite{reviewIC}).

In the following, we will consider 
an extended NJL-type model described by the following Lagrangian:
\begin{eqnarray}
\label{Lagrangianall}
{\cal L} &=& {\bar\psi}\left(\gamma^\mu(i\partial_\mu+\mu\delta_{\mu0}\right)\psi+G\left[({\bar\psi\psi})^2+({\bar\psi i\mathbb{\tau} \gamma_5 \psi})^2\right] 
\nonumber
\\
&+&G'\left[({\bar\psi \sigma^{0j}\psi})^2+ ({\bar\psi i\mathbb{\tau} \gamma_5 \sigma^{0j}\psi})^2\right] \,, 
\end{eqnarray}
 where $\psi$ is the 4$N_f N_c$ dimensional quark spinor with $N_f=2$. Here $\mathbb{\tau}$ are Pauli matrices associated with the SU(2) flavor symmetry and $\sigma^{\nu\rho}=\frac{1}{2}[\gamma^\nu,\gamma^\rho]$.
 
Compared to the conventional NJL model with only scalar and pseudo-scalar channels, the Lagrangian (\ref{Lagrangianall}) has an extra, tensorial channel term with a coupling $G'$. Notice that the two tensor terms are not actually linearly independent, but it is convenient to write them in this particular form.  Just like the standard NJL Lagrangian, Eq.(\ref{Lagrangianall}) 
 is chirally symmetric but does not include confinement and is non-renormalizable.

One could interpret the origin of the different terms of (\ref{Lagrangianall}) as coming from the Fierz transformations of the fermion-gluon QCD vertex. Then one can see that the $G'$ term is zero in vacuum, but appears at finite density due to the explicit breaking of Lorentz symmetry by the chemical potential. The physical value of $G'$ would then depend on the region of densities considered, with $G' \to G$ as the density increases.  

Recall that in quarkyonic matter the main channel of interaction comes from the confining part of the gluon propagator (\ref{confD}). Fierz-transforming the dominant channel leads to
 \begin{equation}
 \label{Fierz}
\gamma^0\gamma^0 = \frac{1}{4} \left\lbrace (\mathbf{1})( \mathbf{1}) + (i \gamma_5)(i\gamma_5)+\sigma^{0i}\sigma^{0i} + ... \,\right\rbrace \,,
\end{equation}
Therefore, in the region of densities and $N_c$ relevant to quarkyonic matter, the scalar and tensor channels have equal strengths. In the following we will then consider the case $G'=G$, keeping in mind that our results will refer to the same region of parameters considered in the case of quarkyonic matter, and hence are not expected to be reliable in describing the low density region.

 Let us now neglect condensation in the pseudoscalar channels and consider a spatially modulated chiral condensate only varying along the z-direction,
$\langle  {\bar\psi}\psi\rangle=S(z)$ and $\langle{\bar\psi}\sigma^{03}\psi\rangle=D(z)$ (see \cite{VFB} for details). Since $\sigma^{03} \propto \gamma^0\gamma^3$, we recognize this as the same kind of ansatz suggested in the context of quarkyonic chiral spirals (cfr. \Eq{sp}). In the mean-field  approximation, the thermodynamic potential is given by
\begin{eqnarray}
\hspace{-.5cm} \Omega(T,\mu; M)&=& 
-\frac{TN_cN_f}{V}\sum_n {\rm Tr}\ln\frac{1}{T}\left[i\omega_n+H_{MF}-\mu \right] 
\nonumber\\ 
&+&\frac{1}{V}\int _{V}\frac{|M(z)|^2}{4G} \,, 
\end{eqnarray}
where the mean-field quark Hamiltonian is given by $H_{MF}=-i\gamma^0\gamma^i\partial_i+\gamma^0\left[ d_{+} M(z)+d_{-}M^*(z) \right]$,
with $d_{\pm}= (1 \pm \gamma^0\gamma^3)/2$ and ${M(z)}=-2G[S(z)+i D(z)]$. The trace acts on Dirac and coordinate space.
The Hamiltonian 
can be block-diagonalized into

\begin{equation}\label{diagonalH}
H^\prime_{MF}=\left(
\begin{array}{cc}
H_{BdG}(M^*(z)) &  (\alpha K)^\dagger \\
(\alpha K) & H_{BdG}(M^*(z))
\end{array}\right) \,,
\end{equation}
where 
\begin{equation}
\label{1D-H}
H_{BdG}(M^*(z))=\left(
\begin{array}{cc}
-i\partial_z & M^*(z)\\
M(z)& i\partial_z
\end{array}\right)
\end{equation}
is the Bogoliubov-DeGennes Hamiltonian and we introduced the 2x2 matrix$$
K = \left(
\begin{array}{cc}
0 & 1 \\
-1 & 0 
\end{array}\right) \,.
$$ 
Performing a Fourier transform in the transverse components, we have  
$\alpha = p_x - i p_y \,,  \vert\alpha\vert^2 = \p_\perp^2 $.

%
In order to determine the eigenvalues of $H_{BdG}(M^*(z))$, we separate the eigenfunction $\psi(z,\mathbf{p}_\perp)$ in two two-component spinors, 
\beq
\psi(z,\mathbf{p}_\perp) = 
\left(
\begin{array}{c}
u(z, \p_\perp) \\ v(z,\p_\perp)
\end{array}
\right) \,.
\eeq
 Then, the eigenvalue equation can be written as a set of two coupled equations:
\beq \label{eigenv-eq}
\left\{\quad {H_{BdG}(M^*) \, u + (\alpha K)^\dagger v = E u}
\atop {H_{BdG}(M^*) \, v + \alpha K u = E v}\right.  \;.
\eeq

We now assume that $u$ and $v$ are eigenfunctions of the $H_{BdG}$ Hamiltonian, characterized by energies $\epsilon(M^*)$ and $\epsilon'(M^*)$ and arrive at 
\bea \label{eq:en3d}
E_\pm(\epsilon,\epsilon',\p_\perp) = \frac{\epsilon(M^*) + \epsilon'(M^*)}{2} \nonumber\\
\pm \frac{1}{2} \sqrt{(\epsilon(M^*) - \epsilon'(M^*))^2 + 4 \p_\perp^2}  \,.
\eea
The relation between $\epsilon$ and $\epsilon'$ can be found starting from \Eq{eigenv-eq}, obtaining
\beq
v = (E - \epsilon) \, \frac{\alpha K}{\vert\alpha\vert^2} \, u  \,.
\eeq
We can now plug this into
\beq 
\epsilon'(M^*) = \frac{v^\dagger H(M^*) v}{v^\dagger v} = 
-\frac{u^\dagger K H(M^*) K u}{u^\dagger u}  \,,
\eeq
and observing that 

\beq
\hspace{-.5cm}
K H(M^*) K =
\left(
 \begin{array}{cc}
-i\partial_z & M\\
M^* & i\partial_z
\end{array}
\right)
= H^T(M^*) = H(M) \,,
\eeq
we arrive at the following relation:
\beq \label{e-relation}
\epsilon'(M^*) = -\epsilon(M) \,.
\eeq
The task of finding the 3+1-dimensional energies $E$ has therefore been reduced to solving a one-dimensional eigenvalue problem 
involving the Bogoliubov-DeGennes Hamiltonian. We note that $H_{BdG}$ is exactly the Hamiltonian of the NJL$_2$ model. 
For the
special case of a plane wave modulation, $M(z)=\Delta e^{iqz}$, the eigenvalues of the Hamiltonian $H(M^*)$ can be obtained by those of $H(M)$ simply by swapping $q$ with $-q$, and the (1+1)-dimensional energies are given by
\beq \label{1dmodes}
\epsilon(M)=\epsilon(\Delta, q)= \frac{q}{2} \pm \sqrt{{p_3}^2+\Delta^2} \,.
\eeq
Plugging them in (\ref{eq:en3d}), one readily finds the spectrum of the (3+1)D theory 
\beq \label{3dmodes}
E(\Delta, q)= -\frac{q}{2} \pm \sqrt{\p^2+\Delta^2} \,,
\eeq
where each modes appears twice.

The first thing we notice is the asymmetry of the 3+1-dimensional energy spectrum, characterized by a $q/2$ 
offset with respect to zero. This is similar to the result obtained for the chiral Gross-Neveu (NJL$_2$) model, but in sharp contrast
with previous studies performed within the standard NJL model (eg. \cite{NickelPRD80,TatsumiPRD71}).
This is due to the different Dirac structure of the condensates in the tensor-extended model considered in the present work: by allowing for condensation in the traditional scalar and pseudoscalar channels, the mean-field Hamiltonian would differ from \Eq{diagonalH} in that instead of $H_{BdG}(M^*)$ one would have $H_{BdG}(M)$ in the upper diagonal block of the matrix. From this, the eigenvalues appear in the thermodynamic potential in a form that resembles more that of the (non-chiral) Gross-Neveu (GN) model, characterized by discrete chiral symmetry (for a more detailed discussion including a Ginzburg-Landau argument, see \cite{NickelPRD80}).

For the case considered in \cite{TatsumiPRD71} of a  chiral density wave oscillating between the scalar and pseudoscalar condensates, $M= -2G(S(z) + i\gamma^5 P(z) )= \Delta e^{iqz}$,
the eigenvalue equations obtained would then be (cfr. \Eq{eigenv-eq} )
\beq
\left\{\quad {H_{BdG}(M) \, u + (\alpha K)^\dagger v = E u}
\atop {H_{BdG}(M^*) \, v + \alpha K u = E v}\right. \,.
\eeq
from which we obtain the relation (cfr. \Eq{eq:en3d})
\bea
\label{eq:en3dsym}
E_\pm(\epsilon,\epsilon',\p_\perp) = \frac{\epsilon(M) + \epsilon'(M^*)}{2} \nonumber\\
\pm \frac{1}{2} \sqrt{(\epsilon(M) - \epsilon'(M^*))^2 + 4 \p_\perp^2}  \,.
\eea
Plugging now the lower-dimensional eigenvalues and using \Eq{e-relation} we obtain
\beq
E_\pm = \pm \sqrt{ \p^2 + M^2 + \bar{q}^2 \pm 2\bar{q} \sqrt{p_z^2 + M^2}} \,,
\eeq
in agreement with the spectrum found in \cite{TatsumiPRD71} with a different method. Here $\bar{q} = q/2$. One can see that the spectrum in this case is symmetric around $E=0$.

\section{Asymmetric Spectrum and Charge Anomaly}
Because of the spectral asymmetry of our effective model, we expect the chiral spiral made of scalar and tensor condensates to be energetically favored in the region of densities where this model is consistent. The reason is that due to the asymmetry of the eigenvalue spectrum, an anomalous contribution to the baryon charge arises in the model thermodynamic potential. This term, proportional to $\mu$ and odd in $q$,
 always decreases the free energy associated to the spiral phase, thus favoring its formation. 

The appearance of an anomalous term dramatically favoring inhomogeneous condensation is not an exclusive feature of the model considered here. Indeed, we know that such kind of term appears naturally in the 1+1-dimensional NJL (or chiral Gross-Neveu) model when considering a plane wave ansatz, and is responsible for the generation of an inhomogeneous phase extending up to arbitrarily high densities \cite{anom1d}. A similar contribution has been shown to appear a in 3+1-dimensional NJL model in presence of an external magnetic field. There the spectral asymmetry is present at the lowest Landau level, and once again can be seen to favor dramatically inhomogeneous phases over homogeneous ones \cite{Spirals-B}.

On the formal side, the arising of such anomalous terms in presence of an asymmetric fermionic spectrum and their contribution to the free energy of the system has been discussed some time ago by Niemi and Semenoff \cite{NS}. Following their recipe, we calculate the anomaly by introducing the quantity $\eta_H$, defined as 
\beq
\eta_H = \lim_{s\to0} \sum_E {\rm sgn}(E)\, \vert E \vert^{-s} \,,
\eeq
where $s$ acts as a regulator for the intermediate steps of the calculation.   
In order to proceed, we 
perform a Mellin transform on the absolute value, giving
\beq
\vert E \vert^{-s} = \frac{1}{\Gamma(s)} \int_0^\infty dw w^{s-1} e^{-\vert E \vert w} \,.
\eeq
In order to obtain an analytical expression for the anomaly, we restrict ourselves to the region $\vert q\vert < \Delta$. There, after 
inserting the expression for our eigenvalues (\Eq{3dmodes}) and summing over the two possible signs in $E$, we have
\bea
\eta_H = \lim_{s\to 0} \, \frac{N_f N_c}{2 \pi^2 \Gamma(s)} \int_0^\infty dw w^{s-1} \left(e^{-q w} - e^{q w}\right)  \nonumber\\
\times \int_0^\infty dp \, p^2 e^{-\sqrt{p^2 + \Delta^2}w} \,,
\eea
and after performing the two integrations and finally taking the $s\to0$ limit, we arrive at 

\beq
\eta_H = - \frac{N_f N_c}{12 \pi^2} \left(2 q^3 - 3 \Delta^2 q \right) \qquad (\vert q\vert < \Delta)\,.
\eeq 

The anomalous term contributes 
to the free energy of the system as 
\beq
\delta\Omega_{anom} = \frac{1}{2} \mu \, \eta_H\,.
\eeq 
From this we can therefore see that as soon as $\mu \neq 0$, the anomalous contribution will favor a nonzero value of $q$, giving immediately rise to the inhomogeneous phase.
This in turn suggests that, in the region of the phase diagram where the model suggested in this work is expected to be realistic (ie. if we can assume $G = G'$ and sufficiently strong coupling), 
the plane wave solution discussed will be energetically favored, consistently with the predictions coming from quarkyonic matter studies.

\section{Conclusions}

While an increasing consensus has been building around the idea that finite-density QCD might be characterized by the formation of a spatially modulated chiral condensate, there are still many open questions on the exact nature and size of the inhomogeneous window.
The present work is a first step towards building a bridge between the quarkyonic matter arguments and a quantitative model study. After implementing a chiral spiral-type solution within an extended NJL model, we obtained an explicit expression for the eigenvalues of the Hamiltonian of the system. The most prominent feature of this type of solution is the asymmetry of the single-particle energy spectrum, which, in analogy with the plane wave case in the one-dimensional chiral Gross-Neveu (NJL$_2$) model, generates an anomalous term which favors inhomogeneous condensation at any nonzero chemical potential. 
It is important to recall that the results discussed rely heavily on the structure of the model, and in particular on the assumption that isoscalar and tensor channel are characterized by the same coupling strength. While, as discussed, this assumption is not expected to be vaild at lower $\mu$, it is naturally justified at higher densities and the results obtained are consistent with the quarkyonic matter predictions, even though some of the fundamental underlying mechanisms such as confinement are presently lacking in our model.
Once having obtained an explicit expression for the energy spectrum of the model as well as some first insight on the phase structure by looking at the anomalous term, the next logical step is to calculate the full phase diagram allowing for the spiral modulation discussed. After having determined the model phase structure, it would be of extreme interest to include the effects of magnetic fields on the inhomogeneous condensate considered. Indeed, it has been shown that an external magnetic field has very interesting consequences on a chiral spiral, the most prominent being the generation of an additional magnetic dipole condensate \cite{ferrer-incera-sanchez}.
These effects could lead to a further enrichment of the already surprisingly variegated phase structure of high-density QCD, and are definitely worth investigating.

\vspace{.2cm} 
{\bf Acknowledgments:} The work of Ferrer and de la Incera has been supported in part by DOE Nuclear Theory grant DE-SC0002179.



%

\nocite{*}
\bibliographystyle{elsarticle-num}
\bibliography{martin}



\end{document}